\documentclass[final]{aipproc}
\layoutstyle{8x11double}
 
\def\xte{{\it RXTE}}

\begin{document}

\title{A study of RXTE and BeppoSAX observations of Cyg X-3}

\author{A. Szostek}{
  address={Centrum Astronomiczne im.\ M. Kopernika, Bartycka 18, 00-716 Warszawa, Poland}
}

\author{A. A. Zdziarski}{
  address={Centrum Astronomiczne im.\ M. Kopernika, Bartycka 18, 00-716 Warszawa, Poland}
}

\begin{abstract} We present an analysis of Cyg X-3 data from {\it RXTE}/PCA, 
HEXTE and ASM, supplemented by a selected spectrum from {\it BeppoSAX}. We fit 
the PCA/HEXTE spectra from 1996--2000 by a model including hybrid 
Comptonization, reflection and absorption, and classify them into hard, 
intermediate and soft states. Apart from the very strong absorption in Cyg X-3, 
the spectra resemble those of GRS 1915+105. The soft and intermediate state 
spectra require the presence of nonthermal Comptonizing electrons. We then study 
the radiative processes at soft X-rays with a hard-state spectrum from {\it 
BeppoSAX\/} modeled including emission from a photoionized plasma. 
\end{abstract}

\maketitle

\section{INTRODUCTION}

Cygnus X-3 is a bright X-ray binary system, located $\sim$9 kpc away 
\citep{Predehl} in the plane of the Galaxy. In spite of its discovery already in 
1966 \citep{giacconi}, it remains poorly understood. The companion is probably a 
Wolf-Rayet star with huge mass loss \citep{keerkwijk}. The nature of the compact 
object remains unknown.

The system exhibits distinct orbital modulation with a 4.8 hr period, observed 
both in X-rays and infrared. In radio, Cyg X-3 is the brightest X-ray binary 
\citep{mccollough99}, and its very strong radio outbursts indicate the presence 
of a jet. In X-rays, it exhibits a wide range of variability patterns. On the 
timescales of months to  years, transitions between the hard and soft spectral 
state occur. The observed lack of ms-timescale variability appears to be due to 
scattering in the strong wind of the companion \citep{mccollough}. 

\section{RXTE SPECTRA}
\label{spectra}

\begin{figure}[ht]\label{individual}
  \includegraphics[width=0.95\columnwidth]{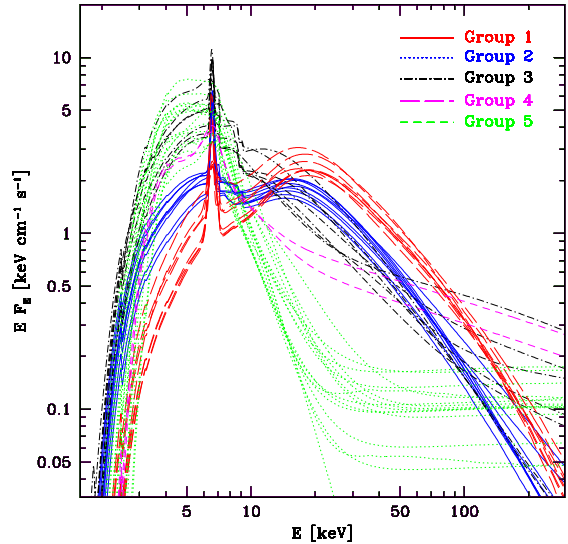}
  \caption{Comptonization model spectra of Cyg X-3 from pointed \xte\/ observations. Different line styles correspond to our classification of the spectra, which show a continuity of spectral shapes. Two extreme states, one with a strong soft X-ray emission followed by a weak hard X-ray tail, and one with a weak soft X-ray emission and hard X-rays peaking around $\sim$20 keV, are clearly seen.}
\end{figure}

We use PCA/HEXTE data from 42 observations in 1996--2000. We use the PCUs 0--2 
and the top layer only. We fit each of the PCA/HEXTE spectra with the same model 
as applied to {\it INTEGRAL}/{\it RXTE\/} spectra from Cyg X-3 \citep{vilhu}. 
The model includes Comptonization by hybrid (i.e., both thermal and nonthermal) 
electrons (\citep{coppi,gierlinski}), Compton reflection from an ionized medium 
\citep{mz95}, and absorption by fully and partially covering neutral media. As 
discussed in \citep{vilhu}, this model treats the low-energy part of the 
spectrum relatively phenomenologically, given the strong absorption taking place 
in Cyg X-3 and the PCA energy coverage limited to $>$3 keV. Still, it provides 
good fits to the PCA/HEXTE spectra and a physical description of the 
hard X-rays. 

\begin{figure}[ht]\label{groups}
  \includegraphics[width=0.98\columnwidth]{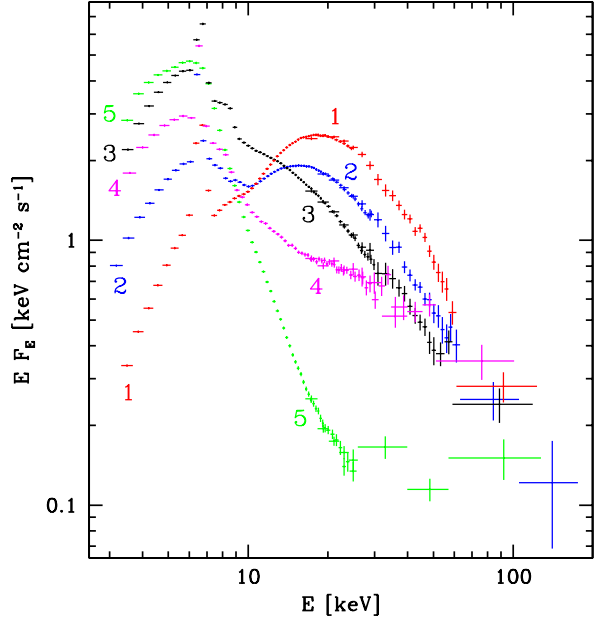}
  \caption{Deconvolved spectra for the average of each of the 5 groups shown on Fig.\ \ref{individual}, fitted with the same model as the individual spectra. The HEXTE data are renormalized to the level of the PCA.
}
\end{figure}

The obtained model spectra are shown in Fig.\ \ref{individual}. We have divided 
them into five groups ordered by the decreasing flux at 20 keV. The increasing 
group number also roughly corresponds to the decreasing spectral hardness in the 
10--20 keV range. We classify the groups 1--2, 3--4, and 5 as belonging to 
the hard, intermediate and soft state, respectively. Some of the flux 
variability within each group is caused by the orbital modulation. 

We have then created average PCA/HEXTE spectra for each group. We have fitted 
them with the same model as above, obtaining $0.3<\chi^2_\nu< 1.9$ (with a 
1\% systematic error). The resulting spectra are shown in Fig.\ \ref{groups}. 

\begin{figure}[ht]\label{asm}
  \includegraphics[width=0.85\columnwidth]{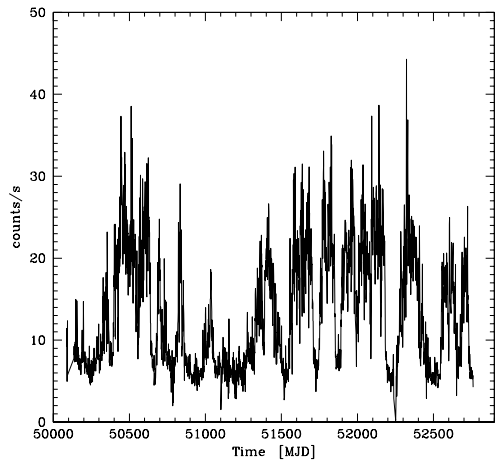}
 \caption{The 1.5--12 keV ASM daily-average lightcurve for MJD 50087--52760. }
\end{figure}

We find that the average spectra require the presence of Comptonizing nonthermal 
electrons to a varying degree. In the hard state, thermal Comptonization appears 
to dominate, and the limited high-energy sensitivity of HEXTE prevents us from 
obtaining precise constraints on nonthermal electrons. The Cyg X-3 spectra 
observed in 2002 Dec.\ by \xte\/ and {\it INTEGRAL\/} \citep{vilhu} 
belong to this state and are closest to our group 2. The intermediate and 
soft-state spectra are dominated by the disk component Comptonized by a 
low-temperature plasma, but then followed by a significant hard tail clearly 
requiring the presence of both hot thermal electrons and nonthermal ones. 

Overall, the high-energy parts of the spectra are quite similar to the spectra 
of GRS 1915+105 shown in \citep{z01}. The hard-state spectra are similar to 
those of the state $\chi$ (in the classification of \citep{belloni}), with the 
hard X-ray photon indices of $\Gamma\simeq 3$. The soft state spectra are then 
similar to the spectrum in the $\gamma$ state of \citep{z01}, with both 
objects showing the characteristic hard tail with $\Gamma\simeq 2$. Note that in 
the case of GRS 1915+105, the presence of nonthermal electrons is required not 
only in the soft state but also in the hard one. The similarity between the two objects represents a very strong argument for the black-hole nature of the compact object in Cyg X-3.

\section{ASM LIGHTCURVES}

\begin{figure}[b]\label{folded}
  \includegraphics[width=0.85\columnwidth]{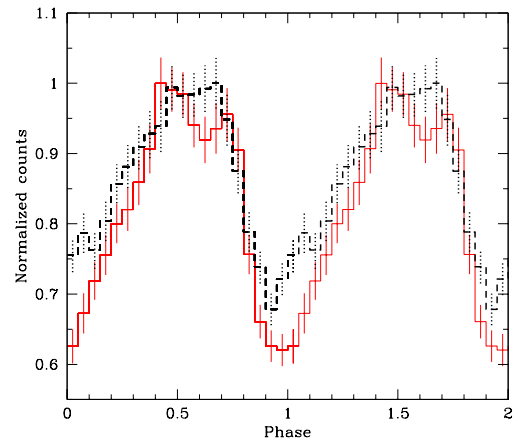}
  \caption{The ASM lightcurves folded with the orbital period for the hard and soft state, plotted in the dashed (black) and solid (red) curve, respectively. 
The period is divided into $20 \times 14.4$ min.\ bins. Each of the lightcurves is normalized to the respective maximum. }
\end{figure}

On the basis of the {\it RXTE}/ASM \citep{levine} lightcurve (Fig.\ \ref{asm}), 
we can clearly distinguish the hard and soft states similar to those defined 
above. Here, somewhat arbitrarily, we define the hard and the soft state by the 
criterion of the countrate smaller and greater than 15 s$^{-1}$, respectively. 
The resulting (dwell-by-dwell) lightcurves have then been folded with the 
orbital period using the parabolic ephemeris \citep{singh}, see Fig.\ 
\ref{folded}.

We find distinct differences in the shape of the two lightcurves. The maximum is broad and flat in the hard state, whereas it forms a sharp peak followed by a secondary one in the soft state. We also see different shapes around the minimum. The differences indicate changes in the source configuration taking place during state transitions. We caution, however, that our criterion to distinguish the states should be phase-dependent as the orbital modulation changes the countrate. Due to this effect, a state with the countrate near 15 s$^{-1}$ gives contribution to either hard or the soft state depending on the phase. We intend to take into account this effect and to study physical implications of the shape of the state-resolved folded lightcurves in the future. 

\begin{figure}[ht] \label{pca_lc}
  \includegraphics[width=0.8\columnwidth]{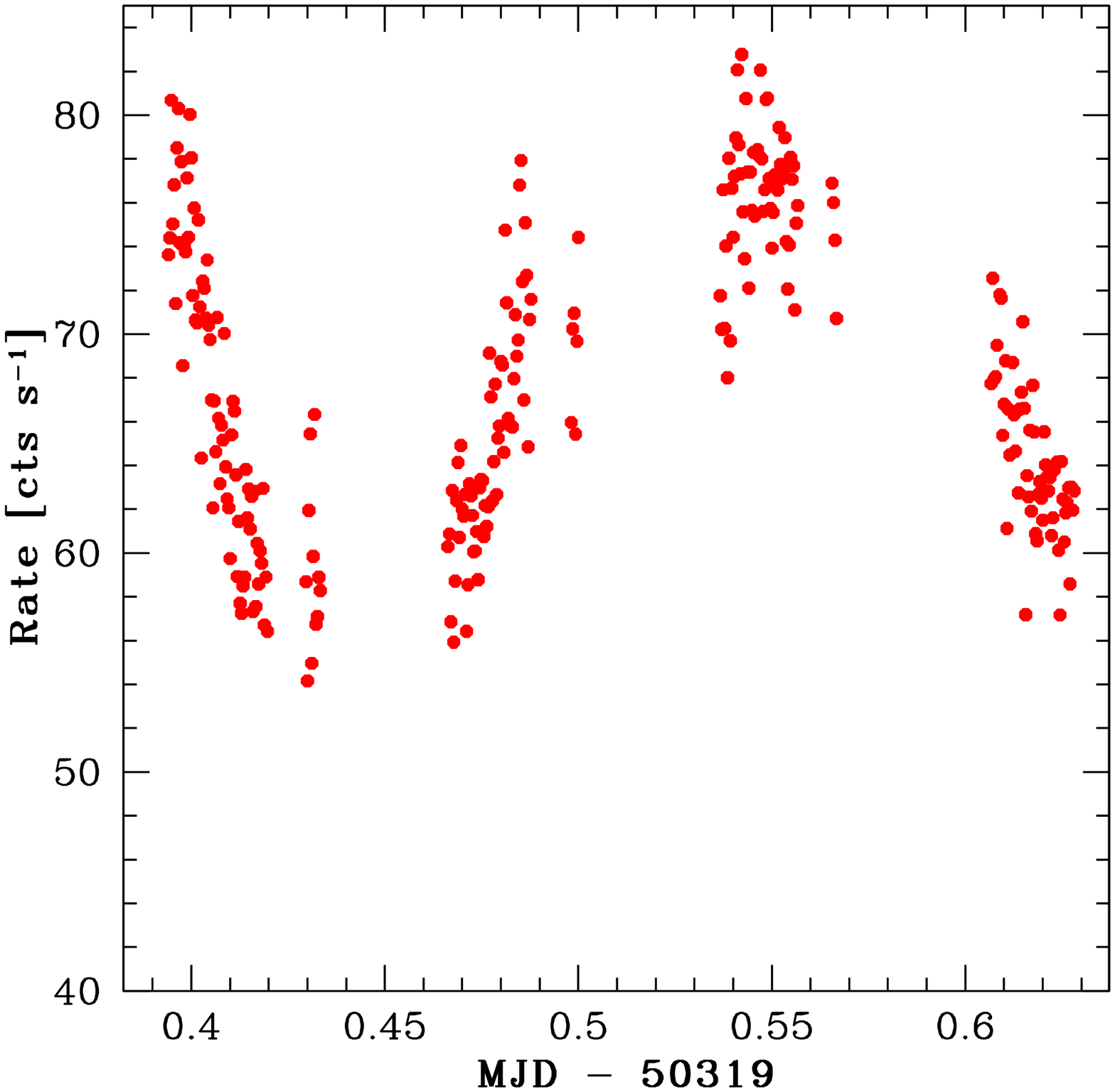}
  \includegraphics[width=0.8\columnwidth]{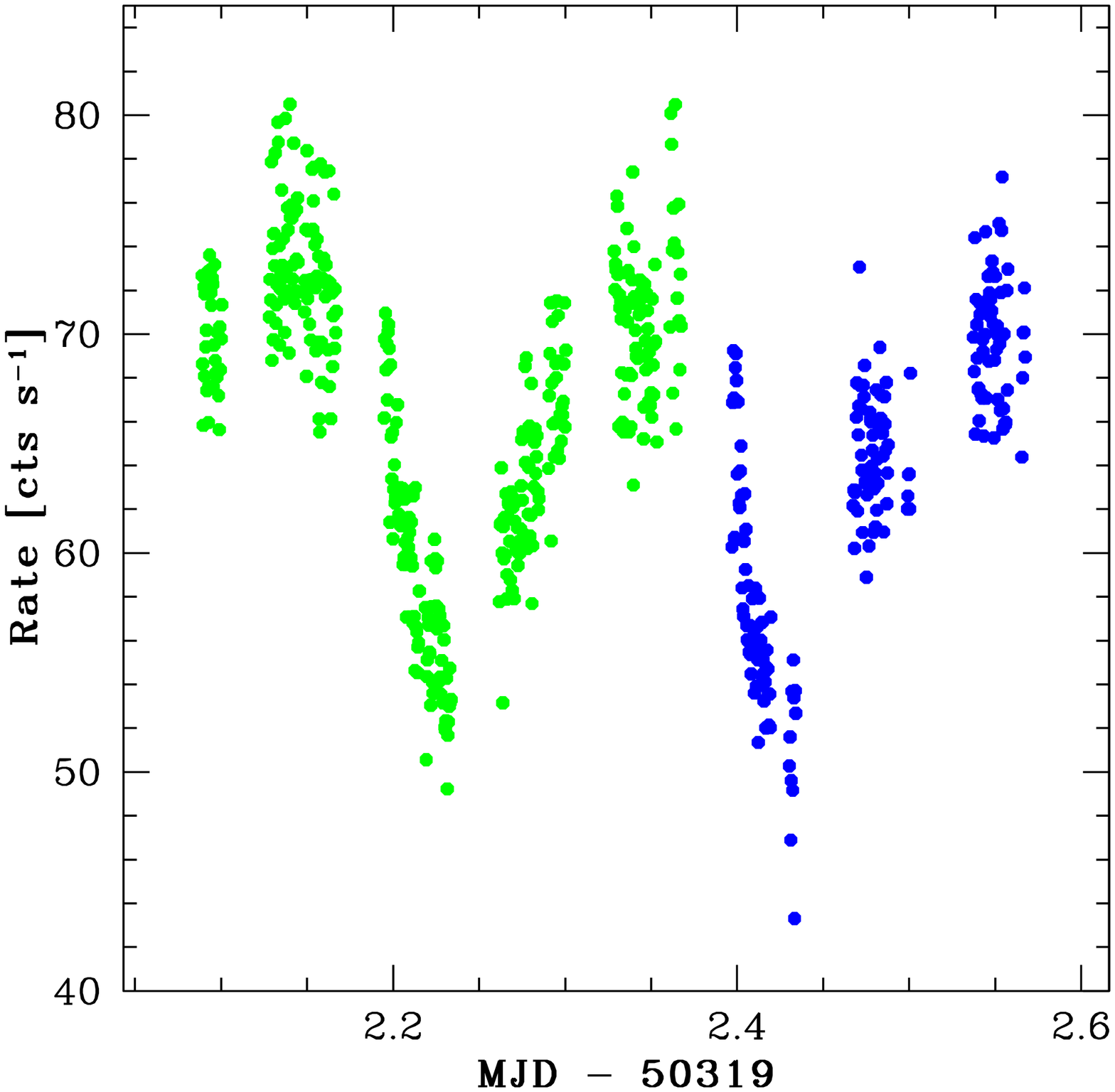}
  \caption{Example PCA lightcurves in the hard state, from observations on 1996 Aug.\ 24  and 26, left and right, respectively. }
\end{figure}

\begin{figure}[ht] \label{phase_spectra}
  \includegraphics[width=0.85\columnwidth]{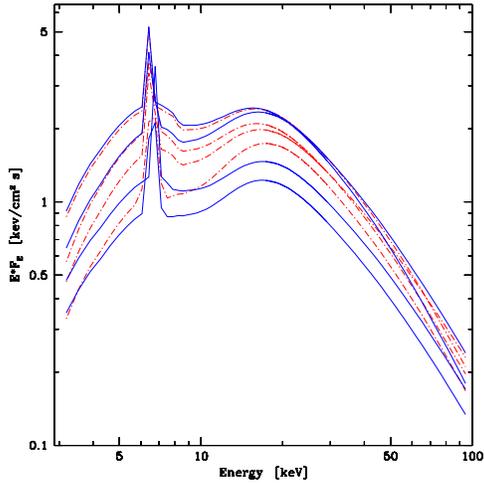}
  \caption{The Comptonization/reflection model fitted to the 8 (out of 18) phase-resolved spectra from Fig.\ \ref{pca_lc}. The dot-dashed (red) curves correspond to the phases of 0--0.5, and the solid (blue) curves, to the phases of 0.5--1.}
\end{figure}

\section{PHASE-RESOLVED RXTE SPECTRA}

We consider now three subsequent \xte\/ pointed observations in the hard state 
in 1996 Aug., see Fig.\ \ref{pca_lc}. We have divided the data into 18 parts, 
each not exceeding in duration 15\% of the orbital period. For each part, we 
have obtained the spectrum, which we then fitted with the model described above. 
We find variability of the overall flux, relatively minor changes in the degree 
of absorption, and distinct changes in the Fe ionization state. Fig.\ 
\ref{phase_spectra} shows 8 selected model spectra.

In the minimum (phase 0/1), there is a low ionization or neutral Fe absorption 
edge at $\sim$7 keV.  On the other hand, a strongly ionized Fe 
absorption edge at $\sim$9 keV is prominent in the maximum (phase 0.5). The edge 
energy and depth indicates the presence of a plasma with He-like Fe with the 
Thomson optical depth $>$1. Due to the low energy resolution of the PCA, it is 
impossible to determine the presence of any narrow absorption or emission 
spectral features. 

\subsection{A FIT TO BeppoSAX DATA}

\begin{figure}[ht] \label{sax}
  \includegraphics[width=0.85\columnwidth]{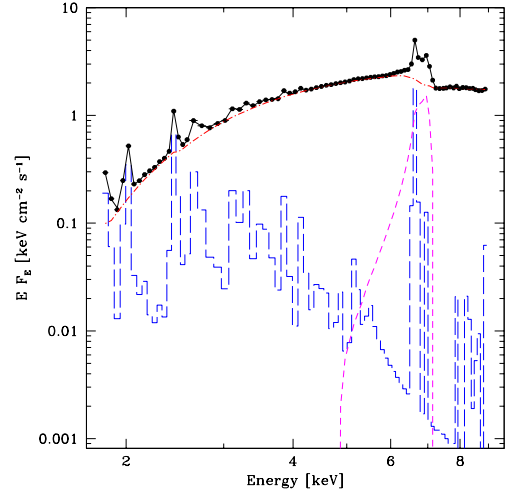}
  \caption{The {\it BeppoSAX}/MECS spectrum (dots/black) fitted by Comptonization/reflection (dot-dashes/red), Fe disk line (short dashes/magenta), emission of an ionized plasma (long dashes/blue), all absorbed by a neutral medium.}
\end{figure}

We then use a hard-state {\it BeppoSAX\/} observation from 1996 Sep.\ 22--24, 
closest in the spectral shape to our group 2 of Figs.\ 
\ref{individual}--\ref{groups}. In this preliminary study, we fit only data from 
the MECS (which has energy resolution much better than that of the PCA), see 
Fig.\ \ref{sax}. We find that the model described above gives a very poor fit 
with $\chi^2_\nu \sim 40$ as it cannot account for the discrete spectral 
features present in the $\sim$1.7--8 keV band. 

Thus, our new model consists of the Comptonization/reflection continuum (as 
above) and a component from a photoionized plasma, all absorbed in neutral 
medium. The absorber includes both the interstellar medium and the wind from the 
companion, with the best-fit total $N_{\rm H}\simeq 3.9\times 10^{22}$ 
cm$^{-2}$. We also include an Fe K$\alpha$ line from an accretion disk 
\citep{fabian}. The emission from the photoionized plasma (calculated using {\tt 
xstar} \citep{kallmann}) turns out to be crucial. It improves the fit 
dramatically, to $\chi^2_\nu\simeq 1.6$. This is due to that component 
accounting for the strong emission features from highly ionized S, Si and Fe 
present in the MECS spectrum. Note that the {\it BeppoSAX\/} observation lasted 
more than nine orbital periods; thus, our results regard the phase-averaged 
spectrum. 

This is our first, and preliminary, attempt of high resolution spectroscopy for 
Cyg X-3. We plan to apply our model to other {\it BeppoSAX\/} observations and 
those by {\it Chandra}.

\section{CONCLUSIONS}

We have obtained several major new results regarding X-rays from Cyg X-3, a 
preliminary account of them is given here. We have obtained the first 
classification of its X-ray spectra, forming 5 classes defined by the spectral 
hardness and the strength of the soft, disk-like component. Each class is well 
fitted by a physical model including Comptonization by both thermal and 
nonthermal electrons and Compton reflection. Apart from the presence of the very 
strong absorption in Cyg X-3, its intrinsic spectra are strikingly similar to 
those of the black-hole binary GRS 1915+105 . This represents a strong argument 
for the presence of a black hole in Cyg X-3. 

In the softest state, we have found a high-energy tail with $\Gamma\simeq 2$ extending above 100 keV. Such a spectral feature also appears in the brightest states of some other black-hole binaries, in particular GRS 1915+105 \citep{z01} and XTE J1550--564 \citep{d02}. The origin of this tail is probably due to Comptonization by nonthermal electrons. 

We then present the \xte/ASM lightcurve of this object, and obtain its profile folded over the orbital period. For the first time, we find distinct differences in its shape between the soft and hard states. This indicates some effects of the different X-ray emission on the surrounding absorbing medium causing the orbital modulation. 

We also investigate the phase-resolved X-ray spectroscopy based on the PCA data. We find the $\sim$9 keV edge from He-like Fe becomes strongest at the maximum, i.e., around the phase 0.5.

Last but not least, we fit the spectrum from an instrument with a medium energy resolution, the {\it BeppoSAX}/MECS. The spectrum, belonging to the hard state, shows a number of {\it emission\/} features, which we fit by a model of photoionized plasma. This component improves the fit to the data from $\chi^2_\nu\simeq 40$ to $\simeq$1.6.  

\begin{theacknowledgments}
This research has been supported in part by KBN grants 5P03D00821, 2P03C00619p1,2, and PBZ-KBN-054/P03/2001. We thank  L. Hjalmarsdotter, P. Lachowicz, F. Paerels, and O. Vilhu for valuable discussions, and the \xte/ASM team for their quick-look results.
\end{theacknowledgments}

\bibliographystyle{aipproc}

\end{document}